 \documentclass[preprint2]{aastex} 

\shorttitle{High Latitude Meridional Flow Variation}
\shortauthors{Rightmire-Upton \&Hathaway }

\begin{document}
\title{Measurements of the Sun's High Latitude Meridional Circulation}

\author{Lisa Rightmire-Upton}
\affil{Department of Physics, The University of Alabama in Huntsville,
Huntsville, AL 35899 USA}
\email{lar0009@uah.edu}

\author{David H. Hathaway}
\affil{NASA Marshall Space Flight Center, Huntsville, AL 35812 USA}
\email{david.hathaway@nasa.gov}

\author{Katie Kosak}
\affil{Florida Institute of Technology, Melbourne, FL 32901 USA}
\email{mkosak2011@my.fit.edu}

\begin{abstract}
The meridional circulation at high latitudes is crucial to the build-up and reversal of the Sun's polar magnetic fields. Here we characterize the axisymmetric flows by applying a magnetic feature cross-correlation procedure to high resolution magnetograms obtained by the Helioseismic and Magnetic Imager (HMI) onboard the Solar Dynamics Observatory (SDO). We focus on Carrington Rotations 2096-2107 (April 2010 to March 2011) -- the overlap interval between HMI and the Michelson Doppler Investigation (MDI). HMI magnetograms averaged over 720 seconds are first mapped into heliographic coordinates. Strips from these maps are then cross-correlated to determine the distances in latitude and longitude that the magnetic element pattern has moved, thus providing meridional flow and differential rotation velocities for each rotation of the Sun. Flow velocities were averaged for the overlap interval and compared to results obtained from MDI data. This comparison indicates that these HMI images are rotated counter-clockwise by $0.075^{\circ}$ with respect to the Sun's rotation axis. The profiles indicate that HMI data can be used to reliably measure these axisymmetric flow velocities to at least within $5^{\circ}$ of the poles. Unlike the noisier MDI measurements, no evidence of a meridional flow counter-cell is seen in either hemisphere with the HMI measurements: poleward flow continues all the way to the poles. Slight North-South asymmetries are observed in the meridional flow. These asymmetries should contribute to the observed asymmetries in the polar fields and the timing of their reversals.

\end{abstract}


\section{INTRODUCTION}

The magnetic fields at the surface of the sun provide the inner boundary condition for the heliosphere. As such, motions of magnetic flux near the solar surface are crucial to studies involving the heliospheric magnetic field configuration, active region evolution, the buildup of polar fields, and subsequent magnetic reversals. Furthermore, surface magnetic flux transport is vital to dynamo models used to explain the sunspot cycle itself. 

The axisymmetric flows were characterized using magnetic feature tracking for nearly the entire MDI era \citep{HathawayRightmire10, HathawayRightmire11}. It was found that the meridional flow varied considerably during that time. Specifically, the meridional flow speed that led up to Solar Cycle 23/24 minimum was much faster than the meridional flow speed during the prior minimum. Faster meridional flow in the active latitudes (equatorward of $\sim 40^{\circ}$) inhibits the cancellation of opposite polarity magnetic elements across the equator. This reduces the imbalance of magnetic polarities in the active latitudes which, when transported to the poles, leads to weaker polar fields, a weaker Solar Cycle 24, and an extended solar minimum between Solar Cycles 23 and 24. While this gives a credible physical explanation for the peculiarities of the Solar Cycle 23/24 minimum, it requires knowledge of the poleward transport all the way to the polar regions. For this reason, as well as others, it is important to measure the meridional flow to the highest latitudes possible.

SDO was launched in February, 2010. On board was HMI \citep{Scherrer_etal12}. HMI is a follow-on and more capable version of MDI. HMI magnetograms, with a size of $4096^{2}$, have four times the spatial resolution of MDI full-disk magnetograms. HMI magnetograms are continuously available with a cadence of 45 seconds, rather than the 96 minutes of MDI. HMI magnetograms averaged over 720 seconds are virtually unaffected by the p-modes \citep{Liu_etal12}. These advances make HMI data ideal for the continuation of the correlation tracking analysis of \cite{HathawayRightmire10, HathawayRightmire11} and ideal for extending the measurements to higher latitudes. 

The meridional flow at high latitudes has significant consequences for models of the Sun's magnetic dynamo. Models of the magnetic flux transport at the Sun's surface \citep{vanBallegooijen_etal98, SchrijverTitle01, Wang_etal09} have employed meridional flow profiles which either stop completely at $75^{\circ}$ latitude or quickly fall to zero before entering the polar regions. Flux Transport Dynamo models depend critically on the strength and structure of the meridional circulation and, with assumptions about the meridional circulation, have been used to predict the amplitude and timing of Solar Cycle 24 \citep{Dikpati_etal06, Choudhuri_etal07}. The presence or absence of counter-cells in the meridional flow at high latitudes has been shown to alter the length of the sunspot cycle in these models \citep{Dikpati_etal10}.

Previous measurements of the meridional motion of the magnetic elements have been limited to lower latitudes. \cite{Komm_etal93} limited their measurements to latitudes less than $52.5^{\circ}$. \cite{Meunier99} included measurement to $70^{\circ}$ while \cite{HathawayRightmire10} stopped at $75^{\circ}$. Measurements of the meridional flow using the methods of helioseismology have been limited to latitudes below $50^{\circ}$ \citep{GonzalezHernandez_etal10, BasuAntia10}. While direct Doppler measurements can conceivably measure the meridional flow right to the poles \citep{Ulrich10}, these measurements are subject to systematic errors from the Convective Blue Shift signal (an apparent blue shift of spectral lines due to the correlation between emergent intensity and radial velocity in convective flows at the surface). This paper seeks to measure the axisymmetric flow velocities, especially the meridional flow, at the highest latitudes possible using magnetic feature tracking with HMI data.

\section{ANALYSIS}

The HMI 720 second magnetogram data were selected for this investigation due to their reduction in noise, particularly near the limb \citep{Liu_etal12}. While spatial resolution is impaired at higher latitudes, this only results in noisier measurements at the higher latitudes (an effect that is mitigated by the high resolution of HMI) without inducing a net flow signal. For this study we chose HMI 720 second full-disk line-of-sight magnetograms obtained every hour on the hour. We account for the radial nature of these magnetic elements by dividing the line-of-sight signal by the cosine of the heliographic angle from disc center as done by \cite{HathawayRightmire10, HathawayRightmire11}. The resulting hourly magnetograms were then mapped into heliographic coordinates, while excluding data within three pixels from the limb. Corrections were made for the ephemeris error in the orientation of the Sun's rotation axis \citep{BeckGiles05, HathawayRightmire10}. Figure 1 shows an example of a mapped image from a 720 second magnetogram.

\begin{figure}[ht!]
\centerline{\includegraphics[width=1.0\columnwidth]{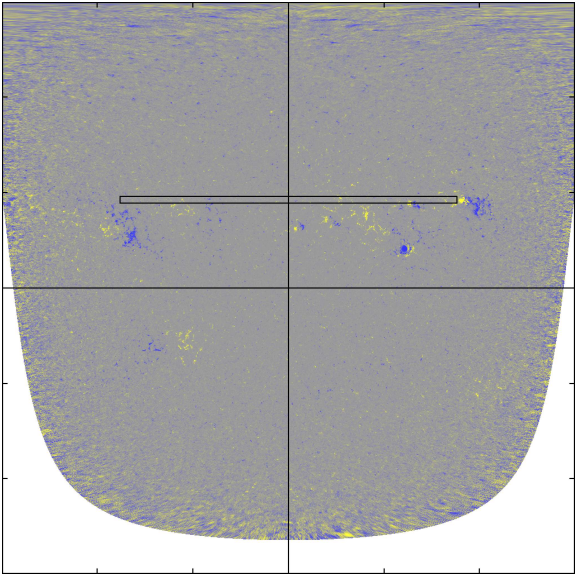}}
\caption{A 720 second averaged magnetogram from 9/2/2010 at 10:00:00 U.T. mapped into heliographic coordinates from pole to pole and ${\pm}90^{\circ}$ from the central meridian and with a resolution of $4096^{2}$. For this date and time the North polar region is well sampled while the South is not. An example of a 41x2401 pixel strip used in the cross-correlation analysis is marked for reference. 
}
\end{figure}

Measurements of the axisymmetric flows require averaging over all longitudes during a full rotation of the Sun. Here Carrington Rotations 2096-2107 were chosen from the $\sim1$ year of overlap between the HMI and MDI observations. Particular attention was paid to the systematic and statistical errors and the consequent latitudinal limits for the two data sources. The measurement method is based on geometry. Errors in the size, shape, or orientation of the images can lead to systematic errors in the desired flows. Using simultaneous measurements from these two instruments allows us to identify these errors. 

The 4096x4096 magnetic maps were divided into 41x2401 pixel strips, one for every  $10^{th}$ pixel latitude position. (One of these strips is shown in Figure 1.) The HMI strips correspond to the same area as the 11x601 pixel MDI strips. In both cases, the latitudinal resolution of the cross-correlations is limited by the strip width to about one half of a degree. Each strip was then cross-correlated with corresponding strips from 8 hours later (no measurements were taken if the shifted strips extended beyond the mapped data). Previous testing \citep{HathawayRightmire11} indicates that this cross-correlation method does in fact measure the axisymmetric motions of the magnetic element pattern. The diffusive effects of the random motions of the magnetic elements due to the nonaxisymmetric convective flows have little or no systematic effects on the measurement results. Measurements were obtained for $\sim 600$ image pairs during each 27-day rotation of the Sun. These measurements were then averaged over the entire rotation and the RMS variations about those averages were calculated. 

The cross-correlation analysis was further improved by incorporating a Forward-Backward technique. In the Forward step (previously used exclusively) a strip centered on the central meridian is taken from the initial map and cross-correlated with shifted strips from the later map. In this case, the displacement is typically  prograde and poleward. In the Backward step, a strip centered on the central meridian is taken from the later map and cross-correlated  with shifted strips from the initial map. In this case, the displacement is typically retrograde and equatorward. The correlations were then combined and the peaks in the subsequent  cross-correlations were fit to parabolas with the position of that peak used to determine flow velocities in both latitude and longitude. By applying this Forward-Backward technique, we are able to double our statistics to further reduce the noise. Moreover, any systematic effect that might mimic flow toward or away from disk center is then canceled in the meridional flow measurements by this Forward-Backward technique.

Measurements were attempted at latitudes up to $85^{\circ}$ in both the North and South but were limited to latitudes less than ${\sim}75^{\circ}$ when the tilt of the Sun's rotation axis was unfavorable. Analysis equatorward of $30^{\circ}$ requires masking out active regions, whose flows are not representative of the surface axisymmetric flows. Pixels with \textbar B\textbar \textgreater 1000 G and the adjacent pixels were masked out in order to exclude flows representative of active regions.

\section{RESULTS}

Axisymmetric flow velocities from each Carrington Rotation were averaged over the Carrington Rotations investigated in this study (2096-2107) using the inverse of the standard errors squared as weights. The HMI average flow velocity profiles were then plotted on top of the average flow velocities obtained using MDI data from the same time periods. Results obtained with HMI have significantly less noise than those from MDI, especially near the poles. 

The HMI meridional flow results were found to be $\sim 2-3$ m s$^{-1}$ more northward than the MDI results. This offset can be explained by a rotation of the HMI imaging system with respect to MDI, as suggested by \cite{Liu_etal12}. (Note that the MDI images were corrected for a $0.19^{\circ}$ rotation relative to the Sun's rotation axis.) Applying a least squares fit to the difference between the two profiles indicates a counter-clockwise rotation of HMI by $0.075^{\circ}$ ( or a velocity correction of ${\sim}2.5 \cos(B)$ m s$^{-1}$, where $B$ is the heliographic latitude). The corrected HMI average flow velocity profiles are plotted along with the  MDI profiles. The differential rotation is shown in Figure 2 and the meridional flow is shown Figure 3. 

\begin{figure}[ht!]
\plotone{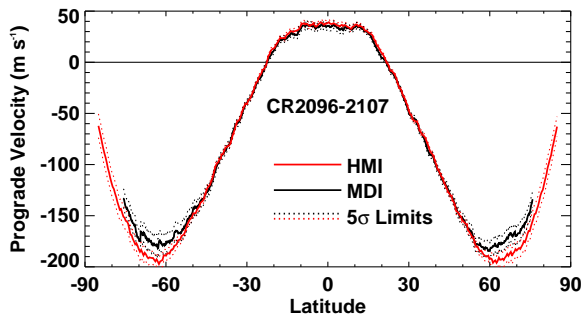}
\caption{The Differential Rotation from Carrington Rotations 2096-2007. The MDI result is plotted in black and the HMI result is plotted in red. The five sigma error ranges are indicated by the dotted lines. 
}
\end{figure}

The MDI and HMI differential rotation equatorward of $55^{\circ}$ are in very good agreement. There are however, small systematic differences ($\sim 1-2$ m s$^{-1}$), with larger velocities in HMI. Poleward of $55^{\circ}$ HMI is $\sim10-20$ m s$^{-1}$ more retrograde relative to the rotating frame of reference than MDI. This may be attributed to a reported elliptical distortion of the MDI images \citep{Korzennik_etal04} which was not completely accounted for in our mapping due to the fact that it is not well characterized. The statistical errors in the HMI data (dotted red lines in Figure 2) indicate that precise measurements can be made to ${\pm}85^{\circ}$ latitude.

\begin{figure}[ht!]
\centerline{\includegraphics[width=1.0\columnwidth]{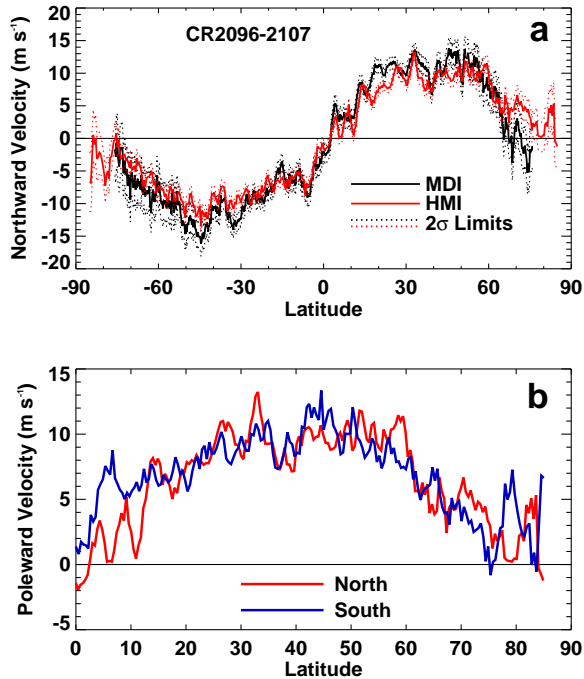}}
\caption{ The Meridional Flow from Carrington Rotations 2096-2007. a) The corrected HMI results are shown in red. Results from MDI are shown in black. The two sigma error ranges are indicated by the dotted lines. b) The poleward velocities in the North (red) and South (blue) are plotted to highlight the North-South asymmetry.}
\end{figure}

HMI measurements of the meridional flow also agree well with those obtained at latitudes up to $\sim 55^\circ$ (Figure 3a), but are systematically slower by $\sim 1-2$ m s$^{-1}$. Again, this can be attributed to a possible elliptical distortion of MDI images. At higher latitudes HMI tells a different story than MDI. The MDI measurements suggested a counter-cell in the North (equatorward flow above $\sim 60^\circ$) but flow to the pole in the South. The HMI measurements have no indication of counter-cells in either hemisphere for this time interval. The results do show a slight, but potentially important, North-South asymmetry. A faster poleward flow is seen in the South through the active latitudes (from the equator to $\sim 40^\circ$) and in the North at higher latitudes. The errors in the HMI meridional flow data (dotted red lines in Figure 3a) indicate that many of fluctuations in the meridional flow profile are actual features, rather than noise. The noise level is low enough that the meridional flow is well determined at $85^\circ$ using HMI data.

\section{CONCLUSIONS}

We find that the improved data from HMI can extend our measurements of the axisymmetric flows, differential rotation and meridional flow, to much higher latitudes -- $85^\circ$ or more. During this analysis, measurements were restricted to $85^\circ$. While noise levels are higher at the poles, it feasible to expect that measurements could be obtained all the way to the pole, particularly for Carrington Rotations in which the sun is tilted $7^\circ$ toward or away from the earth. Our measurements of these flows during the $\sim 1$ year of overlap between the MDI and HMI instrument operations clearly show that the poleward meridional flow extends to the poles. We do not find evidence for any polar counter-cells as was indicated in the highest latitude measurements from MDI.

Recently \cite{Zhao_etal12} found and corrected a systematic error in the meridional flow measurements made with time-distance helioseismology. Their comparison of their corrected meridional flow profile with the contemporaneous meridional flow measurements from magnetic element motions seen with MDI \citep{HathawayRightmire10} and from direct Doppler measurements \citep{Ulrich10} show good agreement to latitudes of $50-60^\circ$ but not much agreement at higher latitudes. The high latitude measurements we show here are largely in agreement with the helioseismology results but further out of line with the direct Doppler measurements (which show counter-cells above $60^\circ$ in each hemisphere). 

We also find slight, but potentially important, North-South asymmetries in the meridional flow profile. The poleward flow is faster in the South in the active latitudes and faster in the North in the polar latitudes. Both of these asymmetries may help to explain the observed North-South asymmetry in the polar fields. \cite{Shiota_etal12} observed a faster decline in magnetic flux of the North pole between 2008 and 2012 than was seen in the South. The trend in the North suggests an imminent polar field reversal. While some of this trend may be due to the fact the the northern hemisphere was more active during this period, the meridional flow asymmetry we observe should also contribute.

The Sun's polar field reversals are produced by the poleward transport of opposite polarity magnetic flux from the active latitudes (opposite to that of the pole at the start of the sunspot cycle). While active regions (sunspot groups) have a balance of both magnetic polarities, in each hemisphere the opposite polarity is systematically found at higher latitudes (Joy's Law). Some of the lower latitude, like polarity, magnetic elements can cross the equator to cancel with similar (but reverse) polarity elements from the other hemisphere and leave behind an excess of the higher latitude, opposite polarity, magnetic elements. A fast meridional flow through the active latitudes inhibits this cross-equator cancellation and leaves a smaller excess of the opposite polarity for subsequent transport to the poles. Thus, the faster meridional flow in the southern active latitudes should result in a slower erosion of the South polar field. In addition, the fast poleward flow at high latitudes in the North should accelerate the erosion of the North polar field.

This poleward flow above $75^\circ$ latitude is, nonetheless, problematic for most models of the magnetic flux transport in the near surface layers \citep{vanBallegooijen_etal98, SchrijverTitle01, Wang_etal09}. In these models  this high latitude poleward flow tends to produce polar magnetic fields that are too strong and too highly concentrated at the poles themselves. Our observations may require further adjustments to those models.

The ultimate consequences of this meridional flow and its asymmetries will require continued measurements of the (variable) meridional flow to high latitudes along with improved models of the flux transport.

\acknowledgements
Hathaway and Rightmire-Upton were supported by a grant from the NASA Living with a Star Program to Marshall Space Flight Center. Kosak was supported by a summer REU Program at The University of Alabama in Huntsville funded by NSF Grant No. AGS-1157027. The HMI data used are courtesy of the NASA/SDO and the HMI science team.

{}

\end{document}